\documentclass[aps,onecolumn,12pt,showpacs,preprintnumbers]{revtex4-1}

\usepackage{amsmath}
\usepackage{bm}
\usepackage{graphicx}
\usepackage{epstopdf}

\begin{document}

\title{Dual axion electrodynamics}

\author{Luca Visinelli}
\email[Electronic address: ]{visinelli@utah.edu}
\affiliation{Department of Physics, University of Utah, 115 S 1400 E $\#$201, Salt Lake City, UT 84102, USA.}
\date{\today}

\begin{abstract}
The duality relation between the electric and magnetic fields, in the presence of an additional axion-like field, is considered. We derive the new equations that describe the electrodynamics in this model, and we discuss the implications for the conservation of the electric and magnetic four-currents and for the propagation of waves.
\end{abstract}


\maketitle

\section{Introduction}

It has long been known that source-free Maxwell's equations for an electro-magnetic (EM) field show an internal symmetry, known as the duality transformation. In a nutshell, 
Maxwell's equations are invariant when the electric and the magnetic fields ${\bf E}$ and ${\bf B}$ mix via a rotation by an arbitrary angle $\xi$, as
\begin{equation} \label{rotation_EM}
\left( \begin{array}{c}
 {\bf E'}\\
 {\bf B'}\\
 \end{array} \right)  =
\left( \begin{array}{cc}
 \cos\xi & \sin\xi\\
 -\sin\xi & \cos\xi\\
 \end{array} \right)\,\left( \begin{array}{c}
 {\bf E}\\
 {\bf B}\\
 \end{array} \right).
\end{equation}
This symmetry can be extended to Maxwell's equations in the presence of sources, provided that additional magnetic charges and currents are introduced in the theory. The duality symmetry has been proven to be associated with a conserved set of currents, as first discussed by Lipkin \cite{lipkin} (see also Refs.~\cite{morgan, kibble, fradkin, karch}).

Whenever a pseudoscalar axion-like field $\theta = \theta(x)$ is introduced in the theory, the dual symmetry is spontaneously and explicitly broken. Axion-like fields and their interactions with the EM field have been intensively studied, see for example Refs.~\cite{sikivie, huang, wilczek, gasperini, jacobs}, and they have recently received attention due to their possible role played in building topological insulators.

Various attempts at constructing an EM Lagrangian that accounts for magnetic sources and axion-like terms have been proposed. Anderson and Arthurs \cite{anderson}, and Rosen \cite{rosen}, proposed a variational technique in which the action is minimized with respect to the field strength $F^{\mu\nu}$, and not the potential $A^\mu$; in Refs.~\cite{anderson, rosen}, the authors obtained the same conserved currents derived in Refs.~\cite{lipkin, morgan, kibble, fradkin}. Sudbery \cite{sudbery} proposed a four-vector Lagrangian $\mathcal{L}_\mu$, invariant under the duality transformation in Eq.~(\ref{rotation_EM}). Using a variational principle for each $\mathcal{L}_\mu$, Sudbery obtained the same results as in Refs.~\cite{anderson, rosen}. Tiwari \cite{tiwari, tiwari1} derived Maxwell's equations in the presence of both electric and magnetic sources, and an axion-like field, imposing the invariance of the local duality symmetry. However, he did not provide a Lagrangian for the system.

In this paper we present a scalar Lagrangian density that accounts for both magnetic sources and the axion-like term, avoiding complications that come from vector Lagrangians. Since both axion-like fields and the duality relations are involved, we call this model the dual axion electrodynamics model. With this Lagrangian, we derive the equation of motion for the fields ${\bf E}$ and ${\bf B}$, corresponding to Maxwell's equations in the presence of electric and magnetic sources, and an axion-like field. In the particular case in which $\theta = 0$ and $\partial_\mu \theta \neq 0$, we obtain the same equations recently obtained by Tiwari \cite{tiwari1}. Finally, we comment on the conservation of the electric and magnetic four-currents and the solution to the source-free set of equations.

To fix our notation, we first review classical electromagnetism in the vacuum and in the presence of an electric charge distribution $\rho_e$ and an electric current ${\bf J}_e$ only.
This paper is organized as follows. In Sec.~\ref{Review of the axion electrodynamics}, we first review Maxwell's formulation of classical EM, and its extension to include a pseudo-scalar axion-like field or magnetic charges and currents (but not both). Sec.~\ref{Dual electrodynamics} is devoted to introducing the dual electrodynamics theory, within which we can write an appropriate Lagrangian for the electrodynamics in the presence of magnetic sources. In Sec.~\ref{Dual axion electrodynamics}, we propose a framework in which the dual electrodynamics theory is modified to accommodate for an axion-like field. In doing this, we present a new EM Lagrangian that accounts for both magnetic sources and an axion-like term, besides the usual electric sources. We also comment on the conservation of the electric and magnetic currents, and the propagation of source-free EM waves.

\section{Review of axion electrodynamics}\label{Review of the axion electrodynamics}

\subsection{Classical electrodynamics}\label{Classical_EM}

In the vacuum, Maxwell's equations for an EM field with electric field ${\bf E}$, magnetic field ${\bf B}$, and the sources $\rho_e$, ${\bf J_e}$ are
\begin{equation}\label{maxwell_equations}
\begin{cases}
{\bf \nabla} \cdot {\bf E} = \rho_e/\epsilon_0,\\
{\bf \nabla} \times {\bf B} -  \epsilon_0\,\mu_0\,\partial_t\,{\bf E} = \mu_0\,{\bf J_e},\\
{\bf \nabla} \cdot {\bf B} = 0,\\
{\bf \nabla} \times {\bf E} + \partial_t\,{\bf B} = 0,
\end{cases}
\end{equation}
where $\partial_t = \partial/\partial t$, and $\epsilon_0$, $\mu_0$ are the electric polarizability and the magnetic permeability of the vacuum, respectively. We remind that the speed of light in the vacuum is given by $c = 1/\sqrt{\epsilon_0\,\mu_0}$. Thanks to the third and fourth lines in Eqs.~(\ref{maxwell_equations}), it is possible to write the electric and magnetic fields in terms of a scalar potential $\phi$ and a vector potential ${\bf A}$, as
\begin{equation}
{\bf E} = -\partial_t\,{\bf A} - {\bf \nabla}\phi,\quad\hbox{and}\quad {\bf B} = {\bf \nabla} \times {\bf A}.
\end{equation}

Maxwell's Eqs.~(\ref{maxwell_equations}) can be reformulated in an elegant form that is explicitly covariant. We first notice that the source $J_e^\mu = (c\,\rho_e, {\bf J_e})$ and the potential $A^{\mu} = (\phi/c, {\bf A})$ transform as four-vectors under a Lorentz transformation. We introduce the antisymmetric field tensor,
\begin{equation}\label{def_F}
F^{\mu\nu} = \partial^\mu A^\nu-\partial^\nu A^\mu,
\end{equation}
in terms of which, the electric and magnetic fields are, respectively
\begin{equation}
E^i = c\,F^{i0} = -\frac{d\phi}{dx^i}-\frac{dA^i}{dt},\quad\hbox{and}\quad B^i = \frac{1}{2}\,\epsilon^{ijk}\,F_{jk} = ({\bf \nabla}\times {\bf A})^i.
\end{equation}
With this notation, Maxwell's equations read
\begin{equation} \label{covariant_eq}
\begin{cases}
\partial_\mu\,F^{\mu\nu} = \mu_0\,J_e^\nu,\\
\partial_\mu\,\tilde{F}^{\mu\nu} = 0.
\end{cases}
\end{equation}
We have introduced the dual of the EM field tensor,
\begin{equation} \label{def_dual}
\tilde{F}^{\mu\nu} = \frac{1}{2}\,\epsilon^{\mu\nu\sigma\rho}\,F_{\sigma\rho}.
\end{equation}
Maxwell's Eqs.~(\ref{covariant_eq}) can be obtained from Euler's equation of motion,
\begin{equation}\label{variational_euler}
\partial^\mu\,\frac{\partial \mathcal{L}_0}{\partial \,(\partial^\mu\,A^\nu)} = \frac{\partial \mathcal{L}_0}{\partial A^\nu},
\end{equation}
using the EM Lagrangian
\begin{equation}\label{Lagrangian_vacuum1}
\mathcal{L}_0 = -\frac{1}{4\,\mu_0}\,F^{\mu\nu}\,F_{\mu\nu} - A_\mu\,J_e^\mu.
\end{equation}

\subsection{Adding an axion-like term}

Maxwell's Eqs.~(\ref{maxwell_equations}) modify if the additional axion-like term
\begin{equation}\label{axion_Lagrangian}
\mathcal{L}_\theta = -\frac{2\,\kappa}{\mu_0\,c}\,\theta\,{\bf E}\cdot{\bf B},
\end{equation}
is added to the Lagrangian in Eq.~(\ref{Lagrangian_vacuum1}). In Eq.~(\ref{axion_Lagrangian}), $\theta = \theta(x)$ is a pseudo-scalar field known in the particle physics literature as the axion-like field, and $\kappa$ is a coupling constant; the factor $1/\mu_0\,c$ assures that $\kappa$ is dimensionless. The effects to Maxwell's equations, of the axion-like term above, have been long discussed, see for example Refs.~\cite{sikivie, huang, wilczek, gasperini, jacobs}. When we consider the EM Lagrangian in Eq.~(\ref{Lagrangian_vacuum1}) with the term in Eq.~(\ref{axion_Lagrangian}) added, the resulting equations of motion for the ${\bf E}$ and ${\bf B}$ fields are Maxwell's Eqs.~(\ref{maxwell_equations}), with the electric charge and current densities replaced by
\begin{equation} \label{substitution}
\rho_e \to \rho_e + \frac{2\kappa}{\mu_0\,c}\,{\bf \nabla}\theta\cdot {\bf B}, \quad \hbox{and}\quad {\bf J_e} \to {\bf J_e} -\frac{2\kappa}{\mu_0\,c}\,({\bf B}\,\partial_t\theta + {\bf \nabla}\theta\times{\bf E}).
\end{equation}
In Ref.~\cite{fedorov}, various physical systems that lead to this type of replacement are discussed. Here, we will just consider the Lagrangian term in Eq.~(\ref{axion_Lagrangian}) for our purposes.

\subsection{Adding magnetic sources} \label{Adding magnetic sources}

Maxwell's Eqs.~(\ref{maxwell_equations}) appear in a more symmetric form when a magnetic charge density $\rho_m$ and a magnetic current ${\bf J_m}$ are included in the theory,
\begin{equation}\label{maxwell_eq_monopole}
\begin{cases}
{\bf \nabla} \cdot {\bf E} = \rho_e/\epsilon_0,\\
{\bf \nabla} \times {\bf B} -  \epsilon_0\,\mu_0\,\partial_t\, {\bf E} = \mu_0\,{\bf J_e},\\
{\bf \nabla} \cdot {\bf B} = \mu_0\,\rho_m,\\
{\bf \nabla} \times {\bf E} + \partial_t\,{\bf B} = -\mu_0\,{\bf J_m}.
\end{cases}
\end{equation}
Similarly to what shown in Eq.~(\ref{covariant_eq}), this last set of equations can be expressed in a covariant form. In fact, defining $J_m^\mu = (c\,\rho_m, {\bf J_m})$, we have
\begin{equation} \label{covariant_eq_monopole}
\begin{cases}
\partial_\mu\,F^{\mu\nu} = \mu_0\,J_e^\nu,\\
\partial_\mu\,\tilde{F}^{\mu\nu} = \mu_0\,J_m^\nu/c.
\end{cases}
\end{equation}
Taking the derivative of the terms in Eqs.~(\ref{covariant_eq_monopole}) with respect to $x^\nu$, we obtain the conservation of the electric and magnetic currents,
\begin{equation} \label{conservation_current}
\begin{cases}
\partial_\nu\,J_e^\nu = 0,\\
\partial_\nu\,J_m^\nu/c = 0.
\end{cases}
\end{equation}
This result is obtained by noticing that $\partial_\mu\partial_\nu$ is a symmetric tensor operator, so that its action on the antisymmetric tensors $F^{\mu\nu}$ and $\tilde{F}^{\mu\nu}$ gives zero.

It can be proven that Eq.~(\ref{covariant_eq_monopole}) is invariant under a rotation by an angle $\xi$,
\begin{equation} \label{rotation_fields}
\left( \begin{array}{c}
 F^{'\mu\nu}\\
 \tilde{F}^{'\mu\nu}\\
 \end{array} \right)  =
U(\xi)\,\left( \begin{array}{c}
 F^{\mu\nu}\\
 \tilde{F}^{\mu\nu}\\
 \end{array} \right),
\end{equation}
when the sources are rotated by
\begin{equation}  \label{rotation_sources}
\left( \begin{array}{c}
 J_e^{'\mu}\\
 J_m^{'\mu}/c\\
 \end{array} \right)  =
U(\xi)\,\left( \begin{array}{c}
 J_e^{\mu}\\
 J_m^{\mu}/c\\
  \end{array} \right).
\end{equation}
Here, we have introduced the SO(2) matrix
\begin{equation}
U(\xi) = \left( \begin{array}{cc}
 \cos\xi & \sin\xi\\
 -\sin\xi & \cos\xi\\
 \end{array} \right).
\end{equation}

\section{Dual electrodynamics} \label{Dual electrodynamics}

To better expose the Lagrangian formulation of electrodynamics in the presence of a electric and magnetic source $J_e^\mu$ and $J_m^\mu$, we introduce a dual four-vector $\tilde{A}^\mu = (\tilde{\phi}/c,\tilde{\bf A})$, so that the dual field strength is defined, similarly to Eq.~(\ref{def_F}) for $F^{\mu\nu}$, as
\begin{equation}\label{def_dualF}
\tilde{F}^{\mu\nu} = \partial^\mu\,\tilde{A}^\nu - \partial^\nu\,\tilde{A}^\mu.
\end{equation}
This approach is not new and it has been considered in early literature (see Refs.~\cite{bisht, baker, cardoso, drummond}).

The relations between the potentials and the electric and magnetic fields now read
\begin{equation}
\begin{cases}
{\bf E} = -{\bf \nabla}\phi - \partial_t\,{\bf A} - {\bf \nabla}\times\tilde{\bf A},\\
{\bf B} = -{\bf \nabla}\tilde{\phi} - \partial_t\,\tilde{\bf A} + {\bf \nabla}\times{\bf A}.
\end{cases}
\end{equation}
Maxwell Eq.~(\ref{covariant_eq_monopole}) can be obtained from the Lagrangian
\begin{equation} \label{Lagrangian_monopole}
\mathcal{L}_{\rm dual} = -\frac{1}{4\,\mu_0}\,F^{\mu\nu}\,F_{\mu\nu}  -\frac{1}{4\,\mu_0}\,\tilde{F}^{\mu\nu}\,\tilde{F}_{\mu\nu} - A_\mu\,J_e^\mu - \tilde{A}_\mu\,J_m^\mu/c,
\end{equation}
when the four-vectors $A^\mu$ and $\tilde{A}^\mu$ are treated independently. We thus have two distinct equations of motion, that resemble Eq.~(\ref{variational_euler}) and lead to
\begin{equation}\label{variational_A}
\partial^\mu\,\frac{\partial \mathcal{L}_{\rm dual}}{\partial \,(\partial^\mu\,A^\nu)} = \frac{\partial \mathcal{L}_{\rm dual}}{\partial A^\nu}, \quad\hbox{and}\quad\partial^\mu\,\frac{\partial \mathcal{L}_{\rm dual}}{\partial \,(\partial^\mu\,\tilde{A}^\nu)} = \frac{\partial \mathcal{L}_{\rm dual}}{\partial \tilde{A}^\nu}.
\end{equation}
From these two variational principles and $\mathcal{L}_{\rm dual}$ in Eq.~(\ref{Lagrangian_monopole}), we re-obtain Maxwell's Eqs.~(\ref{covariant_eq_monopole}).

The dual Lagrangian $\mathcal{L}_{\rm dual}$ is invariant under the dual transformation of the fields in Eq.~(\ref{rotation_fields}), when the four-vectors transform as
\begin{equation}  \label{rotation_vectors}
\left( \begin{array}{c}
 A^{'\mu}\\
 \tilde{A}^{'\mu}\\
 \end{array} \right)  =
U(\xi)\,\left( \begin{array}{c}
 A^\mu\\
 \tilde{A}^\mu\\
  \end{array} \right).
\end{equation}
We now propose a framework to include axion-like terms in the dual Lagrangian in Eq.~(\ref{Lagrangian_monopole}).

\section{The dual axion electrodynamics model} \label{Dual axion electrodynamics}

\subsection{Defining the dual pseudotensor}

Although the Lagrangian in Eq.~(\ref{Lagrangian_monopole}) successfully reproduces electromagnetism with magnetic sources included, it fails when one naively tries to add the axion-like term in Eq.~(\ref{axion_Lagrangian}),
\begin{equation} \label{axion_Lagrangian1}
\mathcal{L}_\theta = -\frac{2\kappa}{\mu_0\,c}\,\theta\,{\bf E}\cdot{\bf B}  = \frac{\kappa}{2\mu_0}\,\theta\,F_{\mu\nu}\,\tilde{F}^{\mu\nu}.
\end{equation}
The main reason why this extra term does not provide the correct equations is that the tensor $\tilde{F}^{\mu\nu}$ has been defined and treated like $F^{\mu\nu}$ and $\tilde{F}^{\mu\nu}$, while in the original EM theory these two tensors are different mathematical objects. In fact, in the EM theory presented in Sec.~\ref{Classical_EM}, we defined $\tilde{F}^{\mu\nu}$ through Eq.~(\ref{def_dual}) so that, if $F^{\mu\nu}$ transforms as a tensor, $\tilde{F}^{\mu\nu}$ transforms as a {\it pseudotensor}. In the dual EM model, in which $\tilde{F}^{\mu\nu}$ is not defined by Eq.~(\ref{def_dual}) but rather by Eq.~(\ref{def_dualF}), the pseudotensorial nature of $\tilde{F}^{\mu\nu}$ is lost in the definition, and must be impose by hand. We thus impose that the contra-variant tensor $\tilde{F}^{\mu\nu}$ and the covariant one $\tilde{F}_{\mu\nu}$ be linked by
\begin{equation} \label{pseudotensor}
\tilde{F}^{\mu\nu} = -\eta^{\mu\sigma}\,\eta^{\nu\rho}\,\tilde{F}_{\sigma\rho}.
\end{equation}
Thanks to this property, the quantity $F^{\mu\nu}\,\tilde{F}_{\mu\nu}$ behaves as a pseudoscalar quantity under a parity transformation. To obtain the transformation of the covariant tensors $F_{\mu\nu}$ and $\tilde{F}_{\mu\nu}$ under a rotation of an angle $\xi$, we use Eq.~(\ref{rotation_fields}) and the fact that $F^{\mu\nu}$  is a tensor and $\tilde{F}^{\mu\nu}$ a pseudo-tensor; we obtain that the covariant tensors transform as
\begin{equation} \label{rotation_fields_covariant}
\left( \begin{array}{c}
 F'_{\mu\nu}\\
 \tilde{F}'_{\mu\nu}\\
 \end{array} \right)  =
U(-\xi)\left( \begin{array}{c}
 F_{\mu\nu}\\
 \tilde{F}_{\mu\nu}\\
 \end{array} \right).
\end{equation}
From the property in Eq.~(\ref{pseudotensor}) it descends the relation
\begin{equation} \label{enforced}
F^{\mu\nu}\,\tilde{F}_{\mu\nu} = -F_{\mu\nu}\,\tilde{F}^{\mu\nu}.
\end{equation}
In the usual EM theory, this property comes from the fact that the Levi-Civita symbol is a pseudotensor, and $\epsilon_{\mu\nu\sigma\rho} = -\epsilon^{\mu\nu\sigma\rho}$. In the following, we enforce Eq.~(\ref{enforced}) in the axion-like Lagrangian term in Eq.~(\ref{axion_Lagrangian1}). The relations
\begin{equation}
F^{\mu\nu}\,F_{\mu\nu} = F_{\mu\nu}\,F^{\mu\nu}, \quad\hbox{and}\quad \tilde{F}^{\mu\nu}\,\tilde{F}_{\mu\nu} = \tilde{F}_{\mu\nu}\,\tilde{F}^{\mu\nu},
\end{equation}
are not modified.

\subsection{The Lagrangian for the system}

It is possible to build a simple Lagrangian for the model, that includes both magnetic sources and the axion-like term, as
\begin{equation} \label{Lagrangian_dual}
\mathcal{L}_{\rm dual+\theta} = -\frac{1}{4\,\mu_0}\,F_{\mu\nu}\,F^{\mu\nu} - \frac{1}{4\,\mu_0}\,\tilde{F}_{\mu\nu}\,\tilde{F}^{\mu\nu} + \frac{\kappa}{2\mu_0}\,\theta\,F_{\mu\nu}\,\tilde{F}^{\mu\nu} - A_\mu\,J_e^\mu - \tilde{A}_\mu\,J_m^\mu/c.
\end{equation}
From the Lagrangian $\mathcal{L}_{\rm dual+\theta}$, we derive the derivatives of the generalized momenta,
\begin{equation} \label{term1.1}
\partial_\mu\,\frac{\partial\mathcal{L}_{\rm dual+\theta}}{\partial\,(\partial_\mu A_\nu)} = -\frac{1}{\mu_0} \,\partial_\mu \left(F^{\mu\nu} - \kappa\,\theta\,\tilde{F}^{\mu\nu}\right),
\end{equation}
\begin{equation} \label{term1.2}
\partial_\mu\,\frac{\partial\mathcal{L}_{\rm dual+\theta}}{\partial\,(\partial_\mu \tilde{A}_\nu)} = -\frac{1}{\mu_0} \,\partial_\mu \left(\tilde{F}^{\mu\nu} + \kappa\,\theta\,F^{\mu\nu}\right).
\end{equation}
The difference in sign in the terms multiplying the axion-like field $\theta$ in Eqs.~(\ref{term1.1}) and~(\ref{term1.2}) comes from enforcing the condition in Eq.~(\ref{enforced}). The derivatives of the Lagrangian with respect to the fields are
\begin{equation} \label{term2.1}
\frac{\partial\mathcal{L}_{\rm dual+\theta}}{\partial\,A_\nu} = - J_e^\nu,
\end{equation}
\begin{equation} \label{term2.2}
\frac{\partial\mathcal{L}_{\rm dual+\theta}}{\partial\,\tilde{A}_\nu} = -J_m^\nu/c.
\end{equation}
Using Eq.~(\ref{variational_A}), we set the term in Eq.~(\ref{term1.1}) equal to the one in Eq.~(\ref{term2.1}), and the term in Eq.~(\ref{term1.2}) equal to the one in Eq.~(\ref{term2.2}). We obtain the field equations
\begin{equation} \label{field_equations}
\begin{cases}
\partial_\mu \left(F^{\mu\nu} - \kappa\,\theta\,\tilde{F}^{\mu\nu}\right) = \mu_0\,J_e^\nu,\\
\partial_\mu \left(\tilde{F}^{\mu\nu} + \kappa\,\theta\,F^{\mu\nu}\right) = \mu_0\,J_m^\nu/c.
\end{cases}
\end{equation}
This set of equations relates the field strengths $F^{\mu\nu}$ and $\tilde{F}^{\mu\nu}$ to the electric and magnetic currents, $J_e^\mu$ and $J_m^\mu$, and to the axion-like field $\theta$. To the best of our knowledge, the set in Eq.~(\ref{field_equations}) has not been previously derived. It is interesting to notice that, since the axion-like term in Eq.~(\ref{axion_Lagrangian1}) explicitly and spontaneously breaks the SO(2) symmetry expressed by the matrix $U(\xi)$, neither the Lagrangian in Eq.~(\ref{Lagrangian_dual}) nor the field Eqs.~(\ref{field_equations}) are invariant under this SO(2) transformation.

We can write the field Eqs.~(\ref{field_equations}) at the CP-preserving configuration $\theta = 0$, with the result
\begin{equation} \label{maxwell_axion}
\begin{cases}
\partial_\mu\,F^{\mu\nu} = \kappa\,\tilde{F}^{\mu\nu}\,\partial_\mu\,\theta + \mu_0\,J_e^\nu,\\
\partial_\mu\,\tilde{F}^{\mu\nu} = -\kappa\,F^{\mu\nu}\,\partial_\mu\,\theta + \mu_0\,J_m^\nu/c.
\end{cases}
\end{equation}
This set of equations was recently obtained by Tiwari in Ref.~\cite{tiwari1}, imposing the conservation of the local form of the SO(2) duality transformation. Notice that at $\theta= 0$, the SO(2) symmetry is restored: the set of Eqs.~(\ref{maxwell_axion}) is indeed invariant under a rotation by an angle $\xi$ of the fields and sources. In the following, we will explore the implications coming from Eqs.~(\ref{maxwell_axion}) to the conservation of the currents and the propagation of EM waves.

\subsection{Propagation of currents and waves}

We now derive the equation for the electric and magnetic currents when an axion-like term is added to the Maxwell Lagrangian. In fact, taking the divergence of Eq.~(\ref{maxwell_axion}), the conservation of currents described by Eqs.~(\ref{conservation_current}) is modified as
\begin{equation}
\begin{cases}
\partial_\mu\,\partial_\nu\,F^{\mu\nu} = \kappa\,\left(\partial_\nu \tilde{F}^{\mu\nu}\right)\,\left(\partial_\mu\,\theta\right) + \kappa\,\tilde{F}^{\mu\nu}\,\left(\partial_\mu\partial_\nu\,\theta\right) + \mu_0\,\partial_\nu\,J_e^\nu,\\
\partial_\mu\,\partial_\nu\,\tilde{F}^{\mu\nu} = -\kappa\,\left(\partial_\nu F^{\mu\nu}\right)\,\left(\partial_\mu\,\theta\right) - \kappa\,F^{\mu\nu}\,\left(\partial_\mu\partial_\nu\,\theta\right) + \mu_0\,\partial_\nu\,J_m^\nu/c,\\
\end{cases}
\end{equation}
or, replacing $\partial_\mu\,F^{\mu\nu}$ and $\partial_\mu\,\tilde{F}^{\mu\nu}$ by means of Eq.~(\ref{maxwell_axion}), 
\begin{equation}\label{maxwell_axion1}
\begin{cases}
\partial_\mu\,\partial_\nu\,F^{\mu\nu} = -\kappa^2\,\left(\partial_\mu\,\theta\right)\,\left(\partial_\mu\,\theta\right)\,F^{\mu\nu} + \kappa\,\tilde{F}^{\mu\nu}\,\left(\partial_\mu\partial_\nu\,\theta\right) + \mu_0\,\left(\partial_\nu\,J_e^\nu-\kappa\,\partial_\mu\theta\,J_m^\mu/c\right),\\
\partial_\mu\,\partial_\nu\,\tilde{F}^{\mu\nu} = -\kappa^2\,\left(\partial_\mu\,\theta\right)\,\left(\partial_\nu\,\theta\right)\,\tilde{F}^{\mu\nu} - \kappa\,F^{\mu\nu}\,\left(\partial_\mu\partial_\nu\,\theta\right) + \mu_0\,\left(\partial_\nu\,J_m^\nu/c+\kappa\,\partial_\mu\theta\,J_e^\mu\right).\\
\end{cases}
\end{equation}
Now, since the tensor operator $\partial_\mu\partial_\nu$ and the tensors $\partial_\mu\theta\,\partial_\nu\theta$ and $\partial_\mu\partial_\nu\theta$ are symmetric, their products with the tensors $F^{\mu\nu}$ and $\tilde{F}^{\mu\nu}$ are zero. Defining
\begin{equation}\label{def_p}
p_\mu = -i \,k\,\partial_\mu\theta,
\end{equation}
Eq.~(\ref{maxwell_axion1}) simplifies as
\begin{equation} \label{current_conservation}
\begin{cases}
\partial_\mu\,J_e^\mu = ip_\mu\,J_m^\mu/c,\\
\partial_\mu\,J_m^\mu/c = -i\,p_\mu\,J_e^\mu.
\end{cases}
\end{equation}
This system of coupled equations is the analogous of Eqs.~(\ref{conservation_current}) in the presence of an axion-like term. The general solution of this system of equations is
\begin{equation} \label{currents_oscill}
\begin{cases}
J_e^\mu = J_+^\mu\,e^{\beta(x)} + J_-^\mu\,e^{-\beta(x)},\\
J_m^\mu/c = -iJ_+^\mu\,e^{\beta(x)} +i J_-^\mu\,e^{-\beta(x)},
\end{cases}
\end{equation}
where $J_+^\mu$ and $J_-^\mu$ are constant four-vectors, and we defined the quantity
\begin{equation}\label{phase}
\beta(x) = \int^x\,p_\sigma\,d{x'}^{\sigma}.
\end{equation}
The system of Eqs.~(\ref{currents_oscill}) describes the damping and the enhancement of the electric and magnetic currents due to the presence of the axion-like field; this effect comes from applying the substitution in Eq.~(\ref{substitution}), for which the axion-like field acts as a source of the electric and magnetic fields. For example, if the currents propagate in space-time, in order to avoid a divergence at infinity we have to impose $J_+^\mu = 0$, so that the terms proportional to $e^{\beta(x)}$ cancel.

We now discuss the propagation of EM waves in the dual axion electrodynamics model. The propagation of EM waves in the axion electrodynamics model has been treated in Ref.~\cite{itin}. The source-free Maxwell's Eqs.~(\ref{maxwell_axion}) for the dual axion electrodynamics model read
\begin{equation} \label{source_free}
\begin{cases}
\partial_\mu\,F^{\mu\nu} = i\,p_\mu\,\tilde{F}^{\mu\nu},\\
\partial_\mu\,\tilde{F}^{\mu\nu} = -i\,p_\mu\,F^{\mu\nu},
\end{cases}
\end{equation}
where we replaced the axion field with $p_\mu$, as in Eq.~(\ref{def_p}). This set of equations looks similar to the one in Eq.~(\ref{currents_oscill}), and has solution
\begin{equation}
\begin{cases}
F^{\mu\nu} = F_+^{\mu\nu}\,e^{\beta(x)} + F_-^{\mu\nu}\,e^{-\beta(x)},\\
\tilde{F}^{\mu\nu} = -iF_+^{\mu\nu}\,e^{\beta(x)} + iF_-^{\mu\nu}\,e^{-\beta(x)},
\end{cases}
\end{equation}
where $F_+^{\mu\nu}$ and $F_-^{\mu\nu}$ are constant, antisymmetric tensors, and $\beta(x)$ is given in Eq.~(\ref{phase}). 

\section{Conclusions}

We have presented a new scalar Lagrangian density, that describes the equations for electrodynamics in the presence of electric and magnetic sources, plus an additional axion-like field $\theta$. Our approach makes use of two vector potential $A^\mu$ and $\tilde{A}^\mu$, describing the fields associated with the electric and the magnetic sources, respectively. We derived a new set of Eqs.~(\ref{field_equations}), that replaces Maxwell's Eqs.~(\ref{covariant_eq}) with these new sources. At the CP-preserving configuration $\theta = 0$, we obtain the same set of Eqs.~(\ref{maxwell_axion}), recently derived by Tiwari \cite{tiwari1}. Finally, we have discussed the implication of this new set of equations to the conservation of currents and the propagation of EM waves.

\end{document}